\documentclass{ws-ijgmmp}

\begin{document}

\markboth{S.Fani, K.Kaviani}
{From Dimentional Reduction ...}

%%%%%%%%%%%%%%%%%%%%% Publisher's Area please ignore %%%%%%%%%%%%%%%
%
\catchline{}{}{}{}{}
%
%%%%%%%%%%%%%%%%%%%%%%%%%%%%%%%%%%%%%%%%%%%%%%%%%%%%%%%%%%%%%%%%%%%%

\title{From Dimensional Reduction of 4d Spin Foam Model to Adding Non-Gravitational Fields to 3d Spin Foam Model%\footnote{For the title, try not to use more than 3 lines. Typeset in 10 pt roman, uppercase and boldface.}
}

\author{Somayeh Fani%\footnote{Typeset names in 8 pt roman, uppercase. Use footnote to indicate permanent address of author.}
}

\address{Physics Department, Alzahra University, Tehran, Iran\,%\footnote{State completely without abbreviations, the affiliation and mailing address, including country.Typeset in 8 pt italic.}
\\
\email{sm.fani@gmail.com%\footnote{Typeset author's e-mail address in 8pt italic}
} }

\author{Kamran Kaviani}

\address{Physics Department, Alzahra University, Tehran, Iran\\
kamran.kaviani@gmail.com }

\maketitle

%\begin{history}
%\received{(Day Month Year)}
%\revised{(Day Month Year)}
%\end{history}

\begin{abstract}
A Kaluza-Klein like approach for a 4d spin foam model is considered. By applying this approach to a model based on group field theory in 4d (TOCY model), and using the Peter-Weyl expansion of the gravitational field, reconstruction of new non gravitational fields and interactions in the action are found. The perturbative expansion of the partition function produces graphs colored with su(2) algebraic data, from which one can reconstruct a 3d simplicial complex representing space-time and its geometry; (like in the Ponzano-Regge formulation of pure 3d quantum gravity), as well as the Feynman graph for typical matter fields. Thus a mechanism for generation of matter and construction of new dimensions are found from pure gravity.
\end{abstract}
\section{Introduction}	
\label{intro}
Spin foam models  \cite{Rovelli-book,Perez,Oriti}, which represents a sum-over-histories approach of quantum gravity, can be defined in any space time dimension. They can be used in different approaches of quantum gravity, such as Loop Quantum Gravity (LQG) (as a path integral formulation for canonical formulation of quantum gravity), topological field theories, and simplicial gravity. Spin foam models are used lately as a general formulation for quantum gravity.
In spin foam models, one may use an abstract 2-complex for illustration of a discrete space time, and assigning algebraic data from the representation of Lorentz group, to it. This combination can assign an algebraic form of geometric data to the simplicial gravity.\\
On the other hand, one can consider spin foam models in terms of the so-called Group field theories (GFT) \cite{Freidel,Oriti4}, i.e. field theories over group manifolds. The GFT formalism \cite{Freidel} represents a generalization of matrix models for 2-dimensional quantum gravity \cite{Francesco6}. There is a group field theory structure for any spin foam model \cite{Pietri7,Reisenberger8}. The GFT structure can be interpreted as a (discrete) third quantization of quantum gravity \cite{Gielen-oriti2011}. Moreover, they allow us  to sum over different topologies \cite{Freidel9,Gurau,Gurau2,Magnen-Noui-Rivasseu-Smerlak,Carroza-Oriti}.\\
 In this picture, spin foams (and thus space-time itself) appear as Feynman diagrams of a field theory which is defined on a group manifold. Thus, spin foam amplitudes are simply the Feynman amplitudes and the number of vertices of the spin foam complex correspond to the orders in perturbative expansion of the GFT.\\
There are some spin foam models, and group field theories, in 4-dimensions, that have been extensively studied \cite{Oriti,newer!!??,newer!!??2}; however their validities are still under investigation. In 4d, one can write a BF theory with some constraints to illustrate 4d gravity. The simplest model to illustrate (topological) 4d gravity without cosmological constant is TOCY\footnote{Turaev, Ooguri, Crane and Yetter }  model \cite{Crane10}, whose group field theory derivation was given by Ooguri \cite{Ooguri11}. In 3-dimentions there are spin foam models of gravity which provide a consistent quantization and are equivalent to those obtained from other approaches. However, each model has some distinctive advantages.  
Indeed, the first model of quantum gravity, the Ponzano-Regge model, was a spin foam model for Euclidean quantum gravity without cosmological constant \cite{Ponzano12}. This approach has been developed to a great extent in the 3-dimensional case. It is now clear that it provides a full quantization of pure gravity \cite{Freidel13}, whose relation with other approaches is well understood \cite{Freidel14,Freidel15}. The presentation of the Ponzano-Regge model as a discrete gauge theory,  was given by Boulatov \cite{Boulatov16}. \\
The above mentioned models only consider pure gravity. There are some recent investigations which  try to couple non-gravitational fields to gravitational field in the framework of spin foam models, such as coupling of the Yang-Mills \cite{Oritti-Pfeiffer,Speziale} and matter fields \cite{Freidel13,Rovelli17,Rovelli18}. This has been done to GFT \cite{Dawdall}as well.\\
In this research the focus is on a Kaluza-Klein like strategy, which can reduce interaction degree in group field theory approach of spin foam model of (topological) 4d-gravity. Also, the symmetry of the theory in 4d is reduced and then as a result, a 3d spin foam model plus the non-gravitational fields is obtained. \\
In section \ref{sec:1}. Ponzano-Regge as a 3d, and TOCY  as a 4d spin foam models are reviewed in framework of GFT. In section \ref{sec:2}. the Kaluza-Klein approach is applied to TOCY model and in the last section the possibility of  how pure gravitational fields can be the source of non-gravitational fields is explained and concluded.
\section{GFT formulation in 3d and 4d}
\label{sec:1}
\subsection{3d-spin foam model (Ponzano-Regge model)}
Based on \cite{Freidel13,Boulatov16} one can consider a real field $\phi(g_{1},g_{2},g_{3})$ 
over a Cartesian product of three copies of $G=SU(2)$ as of the following:
\begin{equation}
\phi:SU(2)\otimes SU(2)\otimes SU(2)\longrightarrow R \label{3d real function}
\end{equation}
It is convenient to require that $\phi$ be invariant under any permutation ($\pi$) of its arguments in the sence that:
\begin{equation}
\phi(g_{1},g_{2},g_{3})=\phi(g_{\pi1},g_{\pi2},g_{\pi3}) \label{3d group field}
\end{equation}
 and invariant under the right diagonal action of SU(2)\footnote{One can impose invariance under diagonal action of the group(G), to give a closure with d number of, (d-2)-faces to form a (d-1)-simplex (in d dimension). (d-1)-simplexes are needed for spin foams.}, in the sence that:
\begin{equation}
 \forall g\in SU(2):\, \, \, \, \, \, \, \, \, \, \, \, \, \, \, \, \, \, \phi(g_{1},g_{2},g_{3})=\phi(g_{1}g,g_{2}g,g_{3}g). \label{3d symmetry1}
\end{equation}
Then it is straightforward to show that:
\begin{equation}
 \phi(g_{1},g_{2},g_{3})=\int_{SU(2)} dh\,\varphi(g_{1}h,g_{2}h,g_{3}h), \label{3d symmetry2}
\end{equation}
where $dh$ is the Haar measure.\\
Now consider a model defined by the following action: 
\begin{eqnarray}
S[\phi]&=&\frac{1}{2}\int\prod_{i=1}^{3}dg_{i}\, \vert\phi(g_{1},g_{2},g_{3})\vert^{2}\nonumber\\
&+&\frac{\lambda}{4!}\int\prod_{i=1}^{6}dg_{i}\, \phi(g_{1},g_{2},g_{3})\phi(g_{3},g_{4},g_{5})\phi(g_{5},g_{2},g_{6})\phi(g_{6},g_{4},g_{1}). \label{3d action}
\end{eqnarray} 
Also, note that any quantum theory like this can be defined by a partition function $Z$,  which can alternatively be expanded in terms of the Feynman diagrams ($\Gamma$) as:
\begin{equation}
Z=\int D\phi\, e^{iS[\phi]} =\sum_{\Gamma}\frac{\lambda^{v}}{sym[\Gamma]} Z(\Gamma),  \label{partition function}
\end{equation}
where $v$ is the number of vertices of the graph $ \Gamma $, $sym[\Gamma]$ is the symmetry factor, and $Z(\Gamma)$ is the Feynman amplitude corresponding to $\Gamma$.\\
It is worth noting that in this case the Feynman diagrams are dual to 3d simplicial complexes, which are supposed to construct the geometrical space. For constructing the corresponding Feynman amplitudes $Z(\Gamma)$, one may consider the following steps:\\
 First going to the representation space, and expanding the field $\phi(g_{1},g_{2},g_{3})$ over $SU(2)\otimes SU(2)\otimes SU(2)$. By using Peter-Weyl theorem, one can write:
\begin{eqnarray}
\phi(g_{1},g_{2},g_{3})=\sum_{j_{1},j_{2},j_{3}}\phi^{j_{1},j_{2},j_{3}}_{m_{1},m_{2},m_{3}}D^{j_{1}}_{m_{1}l_{1}}(g_{1})D^{j_{2}}_{m_{2}l_{2}}(g_{2})D^{j_{3}}_{m_{3}l_{3}}(g_{3})c^{j_{1},j_{2},j_{3}}_{l_{1},l_{2},l_{3}}\label{3d field expansion}
\end{eqnarray}
where $c^{j_{1},j_{2},j_{3}}_{l_{1},l_{2},l_{3}}$ are 3j-symbols\footnote{Clebsh-Gordon coefficients} and $D^{j}_{ml}$ are irreducible representation functions of $g$, and $\phi^{j_{1},j_{2},j_{3}}_{m_{1},m_{2},m_{3}} $ are Fourier like expansion coefficients for $\phi(g_{1},g_{2},g_{3})$.\\
Then, by using (\ref{3d action}) and (\ref{3d field expansion}), one can rewrite the action $S[\phi]$ in terms of the representation space variables in the form of:
\begin{eqnarray}
S[\phi]&=&\frac{1}{2}\sum_{j_{1},j_{2},j_{3}}\vert\phi^{j_{1},j_{2},j_{3}}_{m_{1},m_{2},m_{3}}\vert^{2} %   \nonumber\\
+\frac{\lambda}{4!}\sum_{j_{1},...,j_{6}}\phi^{j_{1},j_{2},j_{3}}_{m_{1},m_{2},m_{3}}\phi^{j_{3},j_{4},j_{5}}_{m_{3},m_{4},m_{5}}\phi^{j_{5},j_{2},j_{6}}_{m_{5},m_{2},m_{6}}\phi^{j_{6},j_{4},j_{1}}_{m_{6},m_{4},m_{1}}\lbrace^{j_{1}\,j_{2}\,j_{3} }_{j_{4}\,j_{5}\,j_{6} }\rbrace, \, \, \, \, \, \, \, \, 
\label{3d expanded action}
\end{eqnarray}\\
where $\lbrace^{j_{1}\,j_{2}\,j_{3} }_{j_{4}\,j_{5}\,j_{6} }\rbrace$ are 6j-symbols\footnote{$
\lbrace 6j \rbrace =\lbrace^{j_{1}\,j_{2}\,j_{3} }_{j_{4}\,j_{5}\,j_{6} }\rbrace=\sum_{l_{1},...,l_{6}} c^{j_{1},j_{2},j_{3}}_{l_{1},l_{2},l_{3}}c^{j_{3},j_{4},j_{5}}_{l_{3},l_{4},l_{5}}c^{j_{5},j_{2},j_{6}}_{l_{5},l_{2},l_{6}}c^{j_{6},j_{4},j_{1}}_{l_{6},l_{4},l_{1}}
$}. \\
Finally, considering (\ref{partition function}) and (\ref{3d expanded action}), one can show the Feynman amplitude of 3d GR \cite{Rovelli-book} as: %$Z({\Gamma})=\sum_{j_{f}}\prod_{f}dim(j_{f})\prod_{v}\lbrace{6j }\rbrace_{v}, $
\begin{equation}
Z({\Gamma})=\sum_{j_{f}}\prod_{f}dim(j_{f})\prod_{v}\lbrace{6j }\rbrace_{v},\label{3d feynman amplitude}
\end{equation}
which is the same as what one may obtain in Ponzano-Regge model.\\
\subsection{4d-spin foam model (TOCY model)}
In 4 dimensions, one can consider a real field $\phi(g_{1},g_{2},g_{3},g_{4})$ over a Cartesian product of four copies of $SO(4)$ which requires to be invariant under the right diagonal action of SO(4). 
Considering the same procedure  as the above 3d GR, the TOCY model \cite{Ooguri11} can be obtained. 
Actually the 4d-action is:
\begin{eqnarray}
S[\phi]=\,\frac{1}{2}\int\prod_{i=1}^{4}dg_{i}&\,& \vert\phi(g_{1},g_{2},g_{3},g_{4})\vert^{2}\nonumber\\
+\frac{\lambda}{5!}\int\prod_{i=1}^{10}dg_{i}&\,& \phi(g_{1},g_{2},g_{3},g_{4})\phi(g_{4},g_{5},g_{6},g_{7})\phi(g_{7},g_{3},g_{8},g_{9})\nonumber\\
 &&\phi(g_{9},g_{6},g_{2},g_{10})\phi(g_{10},g_{8},g_{5},g_{1}).\label{4d action}%  
\end{eqnarray}
 In representation space, using Peter-Weyl decomposition of $\phi$ into unitary irreducible representations, one can write: 
\begin{eqnarray}
\phi(g_{1},g_{2},g_{3},g_{4})=\sum_{j_{1},...,j_{4}}& &\phi^{j_{1},j_{2},j_{3},j_{4},i}_{\alpha_{1},\alpha_{2},\alpha_{3},\alpha_{4}}R^{j_{1}}_{\alpha_{1},\beta_{1}}(g_{1})R^{j_{2}}_{\alpha_{2},\beta_{2}}(g_{2})\nonumber\\
& &R^{j_{3}}_{\alpha_{3},\beta_{3}}(g_{3})R^{j_{4}}_{\alpha_{4},\beta_{4}}(g_{4})v^{i}_{\beta_{1},\beta_{2},\beta_{3},\beta_{4}}, \label{4d expanded field}
\end{eqnarray}
Where $R^{j}_{\alpha,\beta}$ are matrix elements of the unitary irreducible representations $j$ \cite{Rovelli-book}, the indices $\alpha,\beta$ label basis vectors in the corresponding representation space, and the index i represents the orthonormal basis $v^{i}_{\beta_{1},\beta_{2},\beta_{3},\beta_{4}}$ in the space of the intertwiners between the representations $j_{1},j_{2},j_{3},j_{4}$.\\
So, one can rewrite the action in terms of the representation variables as:
\begin{eqnarray}\label{expanded 4d action}
S[\phi]=\frac{1}{2}\sum_{j_{1},j_{2},j_{3},j_{4}}& &\vert\phi^{j_{1},j_{2},j_{3},j_{4}}_{m_{1},m_{2},m_{3},m_{4}}\vert^{2}\nonumber\\
+\frac{\lambda}{5!}\sum_{j_{1},...,j_{10}}& &\phi^{j_{1},j_{2},j_{3},j_{4}}_{m_{1},m_{2},m_{3},m_{4}}\phi^{j_{4},j_{5},j_{6},j_{7}}_{m_{4},m_{5},m_{6},m_{7}}\phi^{j_{7},j_{3},j_{8},j_{9}}_{m_{7},m_{3},m_{8},m_{9}}\nonumber\\
& & \phi^{j_{9},j_{6},j_{2},j_{10}}_{m_{9},m_{6},m_{2},m_{10}}\phi^{j_{10},j_{8},j_{5},j_{1}}_{m_{10},m_{8},m_{5},m_{1}}\lbrace{15j}\rbrace.%\nonumber %\label{expanded 4d action}
\end{eqnarray}
By considering (\ref{partition function}) and (\ref{expanded 4d action}), one can show the Feynman amplitude of (topological) 4d GR (TOCY model) \cite{Rovelli-book} in representation space as %$Z({\Gamma})=\sum_{j_{f},i_{e}}\prod_{f}dim(j_{f})\prod_{v}\lbrace{15j }\rbrace_{v}, $
 of the following:
\begin{equation}
Z({\Gamma})=\sum_{j_{f},i_{e}}\prod_{f}dim(j_{f})\prod_{v}\lbrace{15j }\rbrace_{v},\label{4d feynman amplitude}
\end{equation}
where $ \lbrace{15j }\rbrace$ are the 15j symbols.
\section{Kaluza-Klein strategy}
\label{sec:2}
Returning to the TOCY model, and using SU(2) instead of SO(4)\footnote{BF Theory with gauge group SU(2)\cite{Baez19,Reisenberger18}} for simplicity\footnote{As it is shown in\cite{vilenkin}, any group element g of SO(4) can be written as $g=g^{(3)}h$, where $h\in SO(3)$ and $g^{(3)}$ can be expanded in terms of Euler angels in n dimensional space. If one use $g=g^{(3)}h$ in all calculation presented in this paper and integrate over $g^{(3)}$, it would be expected that, this strategy will work for SO(4) with some relatively complicated calculations!\cite{Fani2}}
, and noting that the action is the same as equation (\ref{4d action}), one can rewrite equation (\ref{4d expanded field}) as:
\begin{equation}
\phi(g_{1},g_{2},g_{3},g_{4})=\sum_{j_{4}}\Phi_{m_{4}l_{4}}^{j_{4}}(g_{1},g_{2},g_{3})D^{j_{4}}_{m_{4}l_{4}}(g_{4}), \label{new expanded 4d field}
\end{equation}
where 
\begin{eqnarray}
\Phi_{m_{4}l_{4}}^{j_{4}}(g_{1},g_{2},g_{3})\equiv \sum_{j_{1},j_{2},j_{3}}\Phi^{j_{1},j_{2},j_{3},j_{4}}_{m_{1}l_{1},...,m_{4}l_{4}}D^{j_{1}}_{m_{1}l_{1}}(g_{1})\,D^{j_{2}}_{m_{2}l_{2}}(g_{2})D^{j_{3}}_{m_{3}l_{3}}(g_{3}) \label{4d auxilary expanded field}
\end{eqnarray}
and
\begin{eqnarray}
\Phi^{j_{1},j_{2},j_{3},j_{4}}_{m_{1}l_{1},...,m_{4}l_{4}}=\sum_{i}\phi^{j_{1},j_{2},j_{3},j_{4},i}_{m_{1},m_{2},m_{3},m_{4}}v^{i}_{l_{1},l_{2},l_{3},l_{4}}. \label{contracted intertwiner}
\end{eqnarray}
Equation (\ref{new expanded 4d field}) is similar to the Fourier expansion of a multi variable function in terms of one of its variables (i.e. same as what is done in Kaluza-Klein approach).\\
 Here, the selected variable for expansion is $g_{4}$; but it is clear that any other variables can be selected as well.\\
Now, by putting (\ref{new expanded 4d field}) in action formula (\ref{4d action}), and considering (\ref{contracted intertwiner}) and integrating over $g_{4}$, one can obtain the following equation for the action:
\begin{eqnarray}
S[\Phi]=\frac{1}{2}\int\prod_{i=1}^{3}dg_{i}\, \sum_{j}&&\frac{\vert \Phi^{j}_{ml}(g_{1},g_{2},g_{3})\vert^{2}}{2j+1}\nonumber\\
+\frac{\lambda'}{4!}\int\prod_{i=1}^{6}dg_{i}\sum_{j_{1},...,j_{4}}&&\frac{\Phi^{j_{1},j_{2},j_{3},j_{4}}_{m_{1}l_{1},...,m_{4}l_{4}}}{(2j_{1}+1)...(2j_{4}+1)}\Phi^{j_{1}}_{m_{1}l_{1}}(g_{1},g_{2},g_{3})\Phi^{j_{2}}_{m_{2}l_{2}}(g_{3},g_{4},g_{5})\nonumber\\
&&\Phi^{j_{3}}_{m_{3}l_{3}}(g_{5},g_{2},g_{6})\Phi^{j_{4}}_{m_{4}l_{4}}(g_{6},g_{4},g_{1}),\label{integrated 4d action}
\end{eqnarray}
It should be noted that, there are some different choices for interaction terms in expanded action. Since close simplexes are considered, the compatible interaction term should be the same as the interaction term in equation (\ref{integrated 4d action}). This means that in action formula (\ref{4d action}), one should expand $g_{1}$ in the first field, $g_{5}$ in the second field, $g_{8}$ in the third field, and $g_{10}$ in the forth field. Then, if close simplexes are considered, one should expand all arguments in the fifth field as well.\\
It is known from LQG that, eigenvalues of area operator are related to j. Therefore, one can see denominators in (\ref{integrated 4d action}) as something proportional to areas.\\
Now one can consider the situation that for every 3-simplexes may assume an area of one of $g_{i}$s is small. Since close simplexes are considered, the result of the previous assumption is that in one of 3-simplexes, all areas become small. This means that in equation (\ref{integrated 4d action}) the areas belong to $g_{1}$(in the 3-simplex that belongs to the first field), $g_{5}$ in the second field, $g_{8}$ in the third field and $g_{10}$ in the forth field are small, and as a result all areas belong to the fifth field will be small. By this requirement, the larger js which belongs to larger areas \cite{Rovelli-book} are inaccessible, and produce smaller terms, and therefore can be regarded as perturbations.\\ 
In other words, looking at integrants in (\ref{integrated 4d action}), one can observe the products of $(2j_{i} + 1)s$ in the denominators, and by assuming that  $\Phi^{j_{1},j_{2},j_{3},j_{4}}_{m_{1}l_{1},...,m_{4}l_{4}}$ are finite, see that the larger js produce the smaller terms. 
Following the strategy of Kaluza-Klein and keeping the $j=0$ term, the action (\ref{integrated 4d action}) will be reduced to:
\begin{eqnarray}
S[\phi]&=&\frac{1}{2}\int\prod_{i=1}^{3}dg_{i}\, \vert \phi(g_{1},g_{2},g_{3})\vert^{2}\nonumber\\
 &+&\frac{\lambda}{4!}\int\prod_{i=1}^{6}dg_{i}\,\phi(g_{1},g_{2},g_{3})\phi(g_{3},g_{4},g_{5})\phi(g_{5},g_{2},g_{6})\phi(g_{6},g_{4},g_{1}) ,
\end{eqnarray}
where the symbol $\phi(g_{1},g_{2},g_{3})$ is substituted for $\Phi^{0}_{00}(g_{1},g_{2},g_{3})$ and $\lambda$ for $\lambda'\Phi^{0000}_{0...0}$.\\
It is known that in quantum level, for producing n-point functions one should integrate over $\Phi^{0000}_{0...0}$. This matter is studied in \ref{app}.\\
This action is the Ponzano-Regge action for 3d gravity with the correct vertex amplitude (i.e., $\lbrace 6j \rbrace $ symbol). As SU(2) is the symmetry group in this study, it is straightforward to show that $\phi(g_{1},g_{2},g_{3})$ has those symmetries that the field in Ponzano-Regge action has.\\
But here, there is a very basic difference in comparison with the Kaluza-Klein approach. Contrary to Kaluza-Klein approach, where the zero mode corresponds to low energy (or large distance) regime, in our approach $j=0$ mode indicates the short distance regime.\\
 For long distance effects, if one lets j become greater than zero, then the first choice is $j=\frac{1}{2}$. On the other hand since in all vertices, the angular momentum is conserved, there will be two choices for interaction terms, which are:\\
1: $j_{1}=j_{2}=\frac{1}{2} $ , $ j_{3}=j_{4}=0$ \,\,\,\,\,\,\,\,and \,\,\,\,\,\,
2. $ j_{1}=j_{2}= j_{3}=j_{4}=\frac{1}{2} $.\\
Using these facts, then the action will be reduced to:
\begin{equation}
S[\phi,\Phi]=S_{3d-pg}[\phi]+S[\phi,\Phi^{\frac{1}{2}}] ,\label{3d pg+ femionik like action}
\end{equation}
where $S_{3d-pg}[\phi]$ is pure gravity action in 3 dimensions, and $S[\phi,\Phi^{\frac{1}{2}}]$ is as follows:\\
\begin{eqnarray}
S[\phi,\Phi^{\frac{1}{2}}]=\frac{1}{2}\int\prod_{i=1}^{3}dg_{i}&&\, \frac{\vert \Phi^{\frac{1}{2}}_{ml}(g_{1},g_{2},g_{3})\vert^{2}}{2}\nonumber\\
+\frac{\lambda'}{4!}\int\prod_{i=1}^{6}dg_{i}&&\, \frac{\Phi^{\frac{1}{2},\frac{1}{2}}_{m_{1}l_{1},m_{2}l_{2}}}{4}\Phi^{\frac{1}{2}}_{m_{1}l_{1}}(g_{1},g_{2},g_{3})\Phi^{\frac{1}{2}}_{m_{2}l_{2}}(g_{3},g_{4},g_{5})\nonumber\\
&&\phi(g_{5},g_{2},g_{6})\phi(g_{6},g_{4},g_{1})\nonumber\\
+\frac{\lambda'}{4!}\int\prod_{i=1}^{6}dg_{i}&&\,\frac{\Phi^{\frac{1}{2},\frac{1}{2},\frac{1}{2},\frac{1}{2}}_{m_{1}l_{1},...,m_{4}l_{4}}}{16}\Phi^{\frac{1}{2}}_{m_{1}l_{1}}(g_{1},g_{2},g_{3}) \Phi^{\frac{1}{2}}_{m_{2}l_{2}}(g_{3},g_{4},g_{5})\nonumber\\
&&\Phi^{\frac{1}{2}}_{m_{3}l_{3}}(g_{5},g_{2},g_{6})\Phi^{\frac{1}{2}}_{m_{4}l_{4}}(g_{6},g_{4},g_{1}) ,\label{fermionik like action}
\end{eqnarray}
As usual, sum over the repeated indices are assumed.\\
 The first term in $S[\phi, \Phi^{\frac{1}{2}}]$ is the kinetic term for $\Phi^{\frac{1}{2}}_{ml}$ and the second term represents how the field $\Phi^{\frac{1}{2}}_{ml}$ interacts with the gravitational field $\phi$. The vertex contribution of this interaction comes from the coupling constant $\frac{\lambda'}{4!}\frac{\Phi^{\frac{1}{2},\frac{1}{2}}_{m_{1}l_{1},m_{2}l_{2}}}{4}$. The last term contains interactions among  $\Phi^{\frac{1}{2}}_{ml}$ fields with coupling constant $\frac{\lambda'}{4!}\frac{\Phi^{\frac{1}{2},\frac{1}{2},\frac{1}{2},\frac{1}{2}}_{m_{1}l_{1},...,m_{4}l_{4}}}{16}$.\\
Now, one can interpret the fields in (\ref{3d pg+ femionik like action}) and (\ref{fermionik like action}) as gravitational and matter like fields, and take $\Phi^{\frac{1}{2}}_{ml}$ as fermionic like fields\footnote{This interpretation can be found in details in \cite{Oriti20}}. By this interpretation, there are three interaction terms in (\ref{3d pg+ femionik like action}). The first one belongs to self interaction of pure gravitational fields, the second one is the interaction of two fermionic like fields with two pure gravitational fields, and the last one is the interaction of four fermionic like fields.\\
If one lets areas become larger, then he/she can include other terms with the $js$ greater than $j=\frac{1}{2}$ in $S[\phi]$. For example, keeping the $j=1$ terms, the action will become:
\begin{equation}
S[\phi,\Phi]=S_{3d-pg}[\phi]+S[\phi,\Phi^{\frac{1}{2}}]+S[\phi,\Phi^{\frac{1}{2}},\Phi^{1}],
\end{equation}
where:
\begin{eqnarray}
S[\phi,\Phi^{\frac{1}{2}},\Phi^{1}]=\frac{1}{2}\int \prod_{i=1}^{3}dg_{i}&&\, \frac{\vert \Phi^{1}_{ml}(g_{1},g_{2},g_{3})\vert^{2}}{3} \, \, \, \, \, \, \, \, \, \, \, \, \, \, \, \, \, \, \, \, \nonumber\\
+\frac{\lambda'}{4!}\int \prod_{i=1}^{6}dg_{i}&&\,\frac{\Phi^{{1},{1}}_{m_{1}l_{1},m_{2}l_{2}}}{9}\Phi^{1}_{m_{1}l_{1}}(g_{1},g_{2},g_{3})\Phi^{1}_{m_{2}l_{2}}(g_{3},g_{4},g_{5})\nonumber\\
&&  \, \,\, \, \, \, \, \, \, \, \, \,\phi(g_{5},g_{2},g_{6})\phi(g_{6},g_{4},g_{1})\nonumber\\
+\frac{\lambda'}{4!}\int \prod_{i=1}^{6}dg_{i}&&\,\frac{\Phi^{\frac{1}{2},\frac{1}{2},1}_{m_{1}l_{1},...,m_{3}l_{3}}}{12}\Phi^{\frac{1}{2}}_{m_{1}l_{1}}(g_{1},g_{2},g_{3})\Phi^{\frac{1}{2}}_{m_{2}l_{2}}(g_{3},g_{4},g_{5})\nonumber\\
&&  \, \, \, \, \,  \Phi^{1}_{m_{3}l_{3}}(g_{5},g_{2},g_{6})\phi(g_{6},g_{4},g_{1})\nonumber\\
+\frac{\lambda'}{4!}\int \prod_{i=1}^{6}dg_{i}&&\,\frac{\Phi^{1,1,1}_{m_{1}l_{1},...,m_{3}l_{3}}}{27}\Phi^{1}_{m_{1}l_{1}}(g_{1},g_{2},g_{3})\Phi^{1}_{m_{2}l_{2}}(g_{3},g_{4},g_{5})\nonumber\\
&&  \, \, \, \, \, \Phi^{1}_{m_{3}l_{3}}(g_{5},g_{2},g_{6})\phi(g_{6},g_{4},g_{1})\nonumber\\
+\frac{\lambda'}{4!}\int \prod_{i=1}^{6}dg_{i}&&\,\frac{\Phi^{1,1,1,1}_{m_{1}l_{1},...,m_{4}l_{4}}}{81}\Phi^{1}_{m_{1}l_{1}}(g_{1},g_{2},g_{3})\Phi^{1}_{m_{2}l_{2}}(g_{3},g_{4},g_{5})\nonumber\\
&& \Phi^{1}_{m_{3}l_{3}}(g_{5},g_{2},g_{6})\Phi^{1}_{m_{4}l_{4}}(g_{6},g_{4},g_{1}) .\label{spin 1 like action}
\end{eqnarray}
This action contains the new fields with new interactions.
 According to \cite{Oriti20} one may interpret the fields $\Phi^{\frac{1}{2}}, \Phi^{1},...,\Phi^{j}$ as different matter fields.\\
 If one takes  $\Phi^{1}$ as a spin 1 like field , the third term in (\ref{spin 1 like action}) is the interaction between two spin $\frac{1}{2}$ like fields and one spin 1 like field; which can be interpreted as the QED\footnote{Quantum electrodynamics}  interaction in ordinary QFT\footnote{Quantum field theory} . 
Now as an option, one may extend the lessons of this calculation to a general picture of higher-$j$ matter fields and their interaction with the 3d gravitational field.
If one starts with a pure gravity in 3 dimensional space and let these dimensions grow up -which is equivalent to appearing the new matter fields in appropriate manner- the new 4 dimensional space which contains a pure 4 dimensional gravity will emerge. In other words, by cooling the 3 dimensional space, many kinds of particles will appear in this space, and finally the appearance of the 4th dimension can be observed. In this new-born space, only a pure 4 dimensional gravity exists.
\section{Conclusions}
The point that can be concluded from the above calculations is that if one starts with a pure gravity in 3 dimensional space and let these dimensions grow up, which is equivalent to appearing the new matter fields in appropriate manner, then what will emerge is the new 4 dimensional space, which contains a pure 4 dimensional gravity. In other words by cooling the 3 dimensional space, many kinds of particles will appear in this space, and finally the appearance of the 4th dimension can be observed. In this new-born space there exists only a pure 4 dimensional gravity.
Actually this is a scenario to show that how the new dimensions and matter (or even dark energy) will appear and extend from pure gravity. 
\appendix
%\section{Appendices}
\section{Partition function}\label{app}
%\label{sec:6}
As it was pointed previously, in quantum theory, to find partition function one should integrate over $\phi^{0000}_{00,00,00,00}$. Let's start from action (\ref{integrated 4d action}), and consider all terms in this action which their $js$ are zero as $S_{0}[\phi]$.
%If one starts from action (\ref{integrated 4d action}), %the terms which in it all js are zerro is as follows:
%the all terms which their js are zeros, are as follows:
\begin{eqnarray}
S_{0}[\phi]=\frac{1}{2}\int\prod_{i=1}^{3}dg_{i}\, \vert \Phi^{0}_{00}(g_{1},g_{2},g_{3})\vert^{2}+\frac{\lambda'}{4!}\int\prod_{i=1}^{6}dg_{i}&&\, \Phi^{0000}_{00,00,00,00}\Phi^{0}_{00}(g_{1},g_{2},g_{3})\Phi^{0}_{00}(g_{3},g_{4},g_{5})\nonumber\\
 && \Phi^{0}_{00}(g_{5},g_{2},g_{6})\Phi^{0}_{00}(g_{6},g_{4},g_{1}) 
\end{eqnarray}
Now, the partition function, $Z_{0}$, can be written as:
\begin{eqnarray}
Z_{0}&=&\int D\phi\, e^{iS_{0}} =\int d\Phi^{0000}_{00,00,00,00}D\phi^{0}_{00} \, e^{iS_{0}} \nonumber\\
&=&\int d\Phi^{0000}_{00,00,00,00} Z_{PR}
\end{eqnarray}
Where $Z_{PR}$ is the partition function of 3-dimensional model, Ponzano-Regge model.
Since $\Phi^{0000}_{00,00,00,00}$ is independent of $g_{i}s$ we use $d\Phi^{0000}_{00,00,00,00}$ instead of $D\Phi^{0000}_{00,00,00,00}$.\\
If one consider the coupling of the theory as $\lambda^{'}\Phi^{0000}_{00,00,00,00}$ and assining it by $\lambda_{0}$, then n-point functions can be written as:
\begin{eqnarray}
\frac{\frac{\partial^{n}Z_{0}}{(\partial j)^{n}}}{Z_{0}}|_{j=0}&=&\int \lambda_{0}(\frac{Z_{PR}}{\int \lambda_{0} Z_{PR}})\frac{\frac{\partial^{n}Z_{PR}}{(\partial j)^{n}}}{Z_{PR}}|_{j=0}
 \label{n point function}
%&=&\int d\Phi^{0000}_{00,00,00,00} \, \rho\frac{\frac{\partial^{n}Z_{PR}}{(\partial j)^{n}}}{Z_{PR}} \label{n point function}
\end{eqnarray}
This is reminiscent of the Kallen-Lehmann theorem \cite{Peskin} in quantum field theory % which is for two point function of the desired theory. 
in which the two point function of interacting theory is written by integrating over two point function of free theory.
\begin{eqnarray}
\frac{\frac{\partial^{2}Z_{int}}{(\partial j)^{2}}}{Z_{int}}|_{j=0}&=&\int \frac{dM^{2}}{2\pi}\rho(M^{2})\frac{\frac{\partial^{2}Z_{free}}{(\partial j)^{2}}}{Z_{free}}|_{j=0} \label{normal n point function}
\end{eqnarray}
where $\rho(M^{2})$ is the weight function.\\
If one sets $(\frac{Z_{PR}}{\int d\Phi^{0000}_{00,00,00,00} Z_{PR}})$ in (\ref{n point function}) as weight function, can see (\ref{n point function}) as generalization of (\ref{normal n point function}) to n-point functions of the theory. %, but here one can write any  n-point functions of interacting theory in terms of .\\
Note that the partition function $Z_{0}$ can be interpreted as partition function of gravity that was reduced from a 4-dimensional space to 3-dimensional one. In this new 3-dimensional space, one can have gravitational field plus other fields. So $Z_{0}$ does not represent pure gravity in 3 dimensional space. \\
If one wants to express partition function, $Z_{0}$, in terms of $Z_{PR}$, which is partition function of pure gravitational action in 3-dimensional space; he/she should integrate over its coupling constant,$\lambda^{'}\phi^{0000}_{00,00,00,00}$.\\
%In other words, as it is known from quantum field theory, writing of any two-point function of interacting theory with respect to free theory, will lead to Kallen-Lehmann theory.  
%Likewise, one can write any n-point functions of \textit{3d gravity plus non-gravitational fields} in terms of n-point functions of  Ponzano-Regge model, which is pure gravity in 3-dimension.\\
%In other words, in quantum field theory, writing of any two-point function of interacting theory with respect to free theory, will lead to Kallen-Lehmann theory. 
%Likewise, one can write any n-point functions of \textit{3d gravity plus non-gravitational fields} in terms of n-point functions of  Ponzano-Regge model, which is pure gravity in 3-dimension.
The above consideration shows that, in quantum level, the partition function of 4-dimensional pure gravity after dimensional reduction from 4 to 3 can be written as the sum of partition functions of pure gravity plus other fields in 3-d with different coupling constant.
%\section*{References}

\end{document}